\documentclass{iopjournal}
\usepackage{ragged2e}
\usepackage[utf8]{inputenc}
\usepackage[T1]{fontenc}
\usepackage{bm}
\usepackage[normalem]{ulem}
\usepackage{lipsum}
\usepackage{graphicx}
\PassOptionsToPackage{table,dvipsnames}{xcolor}
\usepackage{xcolor}
\usepackage{amsmath}
\usepackage{amssymb}
\usepackage{mathtools}
\usepackage{comment}
\usepackage{dcolumn}
\usepackage{hyperref}
\usepackage{soul}
\usepackage{footmisc}
\usepackage{xfrac}
\usepackage{amsfonts}
\usepackage{verbatim}
\usepackage{stackrel}
\usepackage[capitalize]{cleveref}
\usepackage{tikz}
\usepackage{bbold}
\usepackage{float}
\usepackage{physics}

\usetikzlibrary{shapes.geometric, arrows, decorations.markings, decorations.pathmorphing, decorations.pathreplacing}

\begin{document}
\justifying

\articletype{Paper} 

\title{Charging and Discharging a Hubbard–Holstein Quantum Battery: Specific Mechanisms and General Insights}

\author{
Emil \"Ostberg$^{1,*}$
Arvid Steen$^2$
Najmeh Abiri$^3$
Irene D'Amico$^4$
 and Claudio Verdozzi$^{5,*}$
 }

\affil{$^1$Division of Mathematical Physics and ETSF, Lund University, PO Box 118, 221 00 Lund, Sweden}

\affil{$^1$Division of Mathematical Physics and ETSF, Lund University, PO Box 118, 221 00 Lund, Sweden}

\affil{$^3$Department of Computer Science and Media Technology, Malm\"o University, SE-21119 Malmö, Sweden}

\affil{$^4$School of Physics, Engineering and Technology, University of York, York, United Kingdom}

\affil{$^5$Division of Mathematical Physics and ETSF, Lund University, PO Box 118, 221 00 Lund, Sweden}

\affil{$^*$Author to whom any correspondence should be addressed.}

\email{emil.ostberg@fysik.lu.se, claudio.verdozzi@fysik.lu.se}

\keywords{Quantum battery, Hubbard-Holstein, charging, discharging}

\begin{abstract}
\justifying
\noindent A Hubbard--Holstein dimer functions as a correlation-driven quantum battery, 
with ergotropy robustly stored, under some conditions, even in the presence of dissipation. We find 
that, although optimal work extraction can in principle recover all the stored 
energy, it requires unrealistically fine-tuned couplings. By contrast, a 
physically realizable protocol based on spectral matching between the battery 
and the load achieves substantial, albeit suboptimal, energy extraction. Our 
results identify a mechanism for quantum energy storage, provide a realistic 
route to work extraction that is amenable to machine-learning-based theoretical 
exploration, and suggest that quantum batteries may not be universally 
deployable: the microscopic mechanism responsible for storing energy can 
constrain the classes of systems able to efficiently extract it.
\end{abstract}

\section{Introduction}
Advances in technology have enabled experimental platforms to explore thermodynamics at the nanoscale, far from the macroscopic limit, where distinctively quantum resources, such as entanglement and quantum correlations, become available.

In this context, quantum batteries represent a promising paradigm for energy storage \cite{Andolina2019,e26110952,Shaghaghi_2022,Campbell_2026,PhysRevLett.132.210402,Herrera2018,Campaioli2024,GuhaMajumdar2026,PhysRevLett.111.240401,ndlt-qszr,Quach2023}. Unlike conventional batteries, whose charging is constrained by the diffusive transport of charge carriers, quantum batteries can exploit quantum resources such as coherence and many-body interactions, 
as well as bosonic degrees of freedom, including linearly coupled bosonic modes (see, e.g., ~\cite{Downing2023}) and nonlinear bosonic interactions (see, e.g., ~\cite{Andolina2025}).
Taken together, these quantum resources can give rise to a genuine quantum charging advantage \cite{Campaioli2018,Bhattacharjee2021,Campaioli2024,Campbell_2026}.

A wide variety of quantum-battery architectures have been theoretically proposed \cite{Campaioli2024}. Among them, solid-state and solid-state-compatible platforms are especially appealing in view of their potential scalability and integration (see, e.g., Refs.~\cite{Campisi2018,Catalano2024,Horodecki2024}).

However, despite significant progress in charging protocols, the role of phononic excitations and the practical extraction of stored energy remain relatively unexplored in this context. At the same time, some proposed quantum batteries require a continuous energy input to maintain the charged state, calling into question their practical usefulness. Moreover, even when the stored charge is retained, efficient energy extraction under experimentally realistic conditions remains a major challenge. Although ideal work-extraction protocols can, in principle, recover all the stored energy \cite{Allahverdyan_2004}, they generally rely on unphysical couplings and highly engineered interaction Hamiltonians, even for small quantum systems \cite{7pjs-146q}.

In the present work, we consider a two-electron Hubbard--Holstein-type (HH) dimer \cite{Hubbard,Holstein}, containing one spin-up and one spin-down electron, as a minimal model of an electron--phonon quantum battery. In the specific setup considered here, a single local phonon mode couples exclusively to one of the two sites. Although structurally simple, the model captures the interplay of several competing physical mechanisms, including electronic delocalization, Coulomb-driven localization, and phonon-mediated attraction, as well as the associated formation of doublons, polarons, and bipolarons. These mechanisms, and more generally the competition between electron--electron and electron--phonon interactions, have been investigated in both finite and extended Hubbard--Holstein systems (see, e.g., Refs.~\cite{Wellein1996,Jeon2004,Macridin2004,Berciu2007,Zhang2009,Helmer2019}). The dimer therefore provides a minimal setting in which their interplay can be explored. At the same time, it offers a compact framework for investigating aspects of quantum thermodynamics \cite{Emil21,Batge2022,Zhou2024}.

In this context, the electron--phonon coupling in our battery system serves a dual purpose. Besides playing a key role in the charging and discharging dynamics, it provides a physically realistic setting in which to assess the practical limitations of ideal work-extraction protocols. 
Moreover, by coupling the phonon mode to a dissipative bath of harmonic oscillators, we investigate the robustness of the stored energy and ergotropy in the presence of decoherence and dissipation.\footnote{Preliminary aspects of the charging dynamics and of dissipation and ergotropy were investigated in two student papers completed in 2021~\cite{Emil21} and 2026~\cite{Arvid26}, respectively.} The main outcomes of the present work are:

(i) The HH dimer realizes a quantum battery through a storage mechanism involving the interplay between electronic correlations and phononic excitations. The charged state is robust against environmental dissipation for baths with a finite spectral bandwidth.
ii) Formally defined optimal work extraction generally relies on highly fine-tuned, physically unrealistic couplings.
iii) The phonon-induced vibronic spectrum relaxes the spectral-matching requirements between the battery and the load, enabling efficient discharge.
Together, i-iii) identify a virtually unexplored quantum-energy-storage mechanism, clarify the limitations of ideal work-extraction protocols, and establish a physically realizable route to energy extraction, giving  a basis for machine-learning-assisted discharge procedures.

\begin{figure}
\centering
\includegraphics[width=\columnwidth]{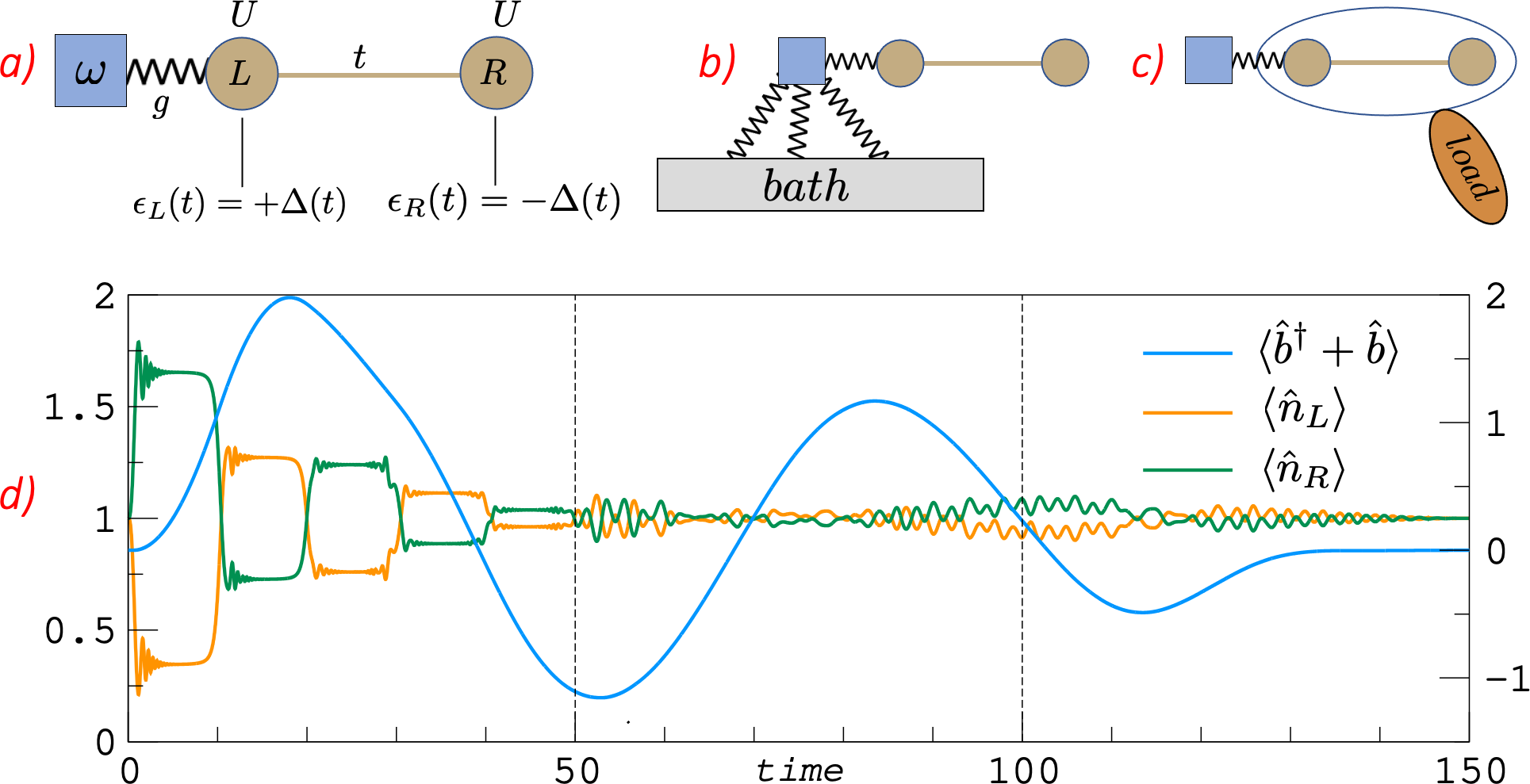}
  \justifying
\caption{Battery schematic and time evolution. (a) Isolated battery (Hubbard-Holstein dimer) and external-field parameters. (b) Battery coupled to a dissipative oscillator bath. (c) Battery--load setup. (d) Time evolution from an initial thermal state ($\beta=2$, $U=2$, $\omega=0.1$, $g=0.2$). Shown are the phonon displacement $\langle \hat{b}^\dagger+\hat{b} \rangle$ and site occupations $\langle \hat{n}_L\rangle$ and $\langle \hat{n}_R\rangle$. Vertical dashed lines separate the charging, dissipation, and discharging regimes.Electronic and phononic quantities use the left and right axes, respectively.}
\label{Fig_one}
\end{figure}
\section{Model and Theoretical Framework}
Our quantum battery is described by the Hamiltonian:
\begin{align}
\hat{H}_{HH} (t)&= \sum_{i\sigma} \epsilon_i (t)\hat{c}_{i\sigma}^\dagger \hat{c}_{i\sigma}
- J \sum_{\langle ij\rangle,\sigma} \hat{c}_{i\sigma}^\dagger \hat{c}_{j\sigma}
+ U \sum_i \hat{n}_{i\uparrow}\hat{n}_{i\downarrow} \nonumber \\
&\quad + \omega \hat{b}^\dagger \hat{b}
+ g(\hat{n}_1 - 1)(\hat{b}^\dagger+\hat{b}), 
\end{align}
where $\hat{c}_{i\sigma}^\dagger$ ($\hat{c}_{i\sigma}$) creates (annihilates) an electron at site $i$ ($=L$ or $R$) with spin $\sigma$ ($=\uparrow$ or $\downarrow$), and $\hat{b}^\dagger$ ($\hat{b}$) creates (annihilates) a phonon. Also, $\hat{n}_{i\sigma} = \hat{c}_{i\sigma}^\dagger \hat{c}_{i\sigma}$ is the electron number operator at site $i$, 
and $\hat{n}_i = \hat{n}_{i\uparrow} + \hat{n}_{i\downarrow}$.

The displaced electronic density  $\hat{n}_L - 1$ at site $L$ is coupled to phonons of frequency $\omega$ with strength $g$. We fix the electronic hopping amplitude  $J$ as the energy unit  ($J=1$). The charging of the battery occurs by applying a time dependent, parity-odd onsite potential to the electronic sites $L,R$, that could physically correspond to an alternating external electric field: $\epsilon_{L}(t) = 
-\epsilon_{R}(t)=\Delta(t)$
(a schematic of the setup is shown in Fig.~\ref{Fig_one}a).
 After the charging, the battery  remains in contact with a bath, to simulate dissipative effects (Fig.~\ref{Fig_one}b). In theoretical quantum-battery research,
dissipative effects are usually introduced via Lindblad dynamics \cite{Manzano,PhysRevApplied}. Here, we adopt a different strategy and 
we consider a bath of $N_\nu$ harmonic oscillators, each coupled linearly to the battery phonon mode. Explicitly,
\begin{align}
\!\!\!\!\hat{H}_{diss.} = \sum_\nu \left[ \frac{\hat{p}_\nu^2}{2} + \frac{1}{2}  \omega_\nu^2 \hat{q}_\nu^2 \right] -\sum_\nu \lambda_\nu \hat{q}_\nu\, (\hat{b}^\dagger+\hat{b}), \label{CLdiss}
\end{align}
which represents the dissipative environment within a Caldeira-Leggett (CL) framework without counter-term proportional to $(\hat{b}^\dagger+\hat{b})^2$ \cite{SciPost,PhysRevE.98.012122,CALDEIRA1983374}.
The battery-bath coupling is treated at the mean-field level, resulting
in the coupled equations of motion
\begin{subequations}
\label{MFEA}
\begin{align}
\frac{d\hat{\rho}}{dt} &= [\hat{H}_{HH}(t) -\sum_\nu \lambda_\nu q_\nu(t)\, (\hat{b}^\dagger+\hat{b}), \hat{\rho}], \label{MFEA1}\\
\ddot{q}_\nu(t) &= - \omega_\nu^2 q_\nu(t)+\lambda_\nu \langle \hat{b}^\dagger+\hat{b} \rangle \label{MFEA2}.
\end{align}
\end{subequations}
where $\hat{\rho}$ is the density matrix of the HH dimer, and $q_\nu(t)\equiv\langle \hat{q}_\nu\rangle_t$. We solve Eq.~(\ref{MFEA1}) numerically using a Suzuki--Trotter decomposition with exact diagonalization (see App.~\ref{Appendice3}), while the bath coordinates in Eq.~(\ref{MFEA2}) are propagated with the Verlet algorithm. Numerically, the density matrix $\hat{\rho}$ has dimension $4N_{ph}$, corresponding to four electronic and $N_{ph}=50$ phononic states.

Since the bath is harmonic and the battery--bath coupling is linear, the decoupling of Eqs.~(\ref{MFEA1},\ref{MFEA2}) corresponds to the mixed quantum--classical Ehrenfest approximation. The latter does not, in general, satisfy detailed balance \cite{DetBal1,DetBal2,DetBal3,DetBal4,DetBal5} and, owing to the absence of stochastic transitions, typically overestimates quantum coherence.
By contrast, standard Markovian Lindblad descriptions integrate out the bath degrees of freedom, replacing the explicit bath dynamics with a spectrally featureless dissipative kernel that couples all energies equally, effectively corresponding to an infinitely broad spectrum. In the present treatment, dissipation arises from a large collection of harmonic modes linearly coupled to the battery phonon, making the Ehrenfest approximation particularly well suited to capture the dissipative dynamics qualitatively, with deviations from the exact, full quantum solution expected to remain primarily quantitative.

As a key observable in characterizing the performance of the battery, we use the ergotropy, i.e. the maximum extractable work $\mathcal{W}$, defined at any time $t$ as \cite{Allahverdyan_2004}
\begin{equation}
\!\!\!\mathcal{W}(\rho(t),H(t)) = \sum_{ij} \epsilon_i (t)r_j (t)\left( |\langle \epsilon_i (t)| r_j (t)\rangle|^2 - \delta_{ij} \right),
\end{equation}
where $\rho(t)\lvert r_j (t)\rangle = r_j(t) \lvert r_j (t)\rangle$ and $H(t)\lvert \epsilon_i(t) \rangle= \epsilon_i(t)\lvert \epsilon_i (t)\rangle_t$. 
When starting from a passive state \cite{Pusz1978,Lenard1978,Allahverdyan_2004} and in the absence of dissipation,
\begin{equation}
\mathcal{W}(t) = \mathrm{Tr}\left[H_{HH}(t)\rho(t)\right] - \mathrm{Tr}\left[H_{HH}(0)\rho(0)\right],
\end{equation}
providing a direct measure of the extractable energy. The results that follow will address the different stages of battery operation: charging, charge retention in the presence of dissipation, and discharging, with an in-depth discussion
of the latter.

\section {Battery dynamics vs robustness to dissipation.}
Initially, the battery is in a thermal state and coupled to the CL bath, with $\beta=2$ and $\ev{\hat{x}}=0$ at $t=0$, implying that the HH dimer is effectively decoupled from the oscillators. We use $N_\nu=1000$, $\omega_\nu=\nu \Delta$, with $\nu=1,\hdots,N_\nu$, $\Delta=0.001$, $m_\nu=1$, $\lambda_\nu = A\omega_\nu^a$, $a=0.2$ and $A=9\cdot 10^{-4}$.
These bath parameters are chosen to ensure a finite energy bandwidth and  not too strong dissipation, together with finite memory effects characteristic of a sub-Ohmic spectral density (a single harmonic oscillator coupled to this type of bath exhibits damped oscillations with an approximately exponential envelope and weak bath-induced revivals). The battery is then charged by applying an external field
$
\epsilon_L(t)=-\epsilon_R(t)=\Theta(T_{\rm max}-t)\,\xi\sin(\pi t/\xi),
$
with $T_{\rm max}=50$ and $\xi=10$, and $\Theta$ the Heaviside function.  The field is then switched off to probe charge retention, allowing the system to evolve freely while still coupled to the bath, before the battery is discharged using the protocol of Ref.~\cite{Allahverdyan_2004}, discussed in detail later in the paper.

Figure~\ref{Fig_one} shows the electronic densities $\langle \hat n_L \rangle, \langle \hat n_R \rangle$ and the average phonon displacement $\langle \hat b^\dagger + \hat b\rangle$ during the three stages (separated by vertical black lines). 
During charging, the field first induces a strong charge imbalance, after which the phonons absorb energy while the electronic densities re-equilibrate. More in detail, the electronic occupations $\langle \hat n_L\rangle$ and $\langle \hat n_R\rangle$, initially equal, develop a pronounced imbalance within the first half-cycle of the perturbation. They subsequently form quasi-plateaus with superimposed high-frequency oscillations, arising from the competition between the onsite-energy imbalance and intersite hopping, and rapidly exchange their relative magnitudes near the boundaries between consecutive half-cycles, where $\Delta\simeq0$. The difference between the occupations decreases after each exchange and it is rather small by the end of the charging protocol. After the charging ends, the electronic densities remain close to half filling, while the phonon coordinate exhibits damped oscillations. During discharging, the electronic densities settle at half filling and $\langle b^\dagger+b\rangle$ decays to zero.

Figure~\ref{Fig_two} shows the corresponding ergotropy, $\mathcal{W}(t)$, which increases during charging, remains nearly constant after the drive is switched off, and eventually decays to zero during discharging. 
In contrast to this alternating electronic dynamics of Fig.~\ref{Fig_one}, 
$\mathcal{W}(t)$ exhibits a cumulative step-like increase through progressively higher quasi-plateaus, whose duration is likewise set by the half-cycles of the perturbation. The oscillations within these plateaus occur on approximately the same timescale as those of the electronic occupations. Meanwhile, the average phonon number, $n_{\rm ph}\equiv\langle b^\dagger b\rangle$, increases overall and displays distinct peaks near the transitions between successive ergotropy plateaus. 
This behavior is consistent with the phonon mode acting as an energy reservoir, storing energy during charging and releasing it slowly to the bath, thereby enhancing the robustness of the ergotropy against dissipation. Related charging and dissipation effects associated with non-thermal initial states were reported in Refs.~\cite{Emil21,Arvid26}.

We now consider the purely dissipative stage, $50\leq t\leq100$. Both $\mathcal{W}(t)$ and $n_{\rm ph}$ remain nearly constant, apart from small variations associated with weak residual dissipation, while the oscillations of $\langle b^\dagger +b\rangle$ are only partially damped. The finite bandwidth of the bath plays an important role in this behavior, a point that will be discussed in detail below in Sect.~\ref{nonoptprot}.
\section{Discharging the Battery: optimal protocol} 
Having seen that  in our setup the battery charges and retains the charge, we address in detail the discharging phase. As established in Ref.~\cite{Allahverdyan_2004}, a closed quantum battery can always be fully discharged. Denoting the final discharge time by $\tau$, this is achieved by constructing the unitary operator $\hat U(\tau)=\sum_k |\epsilon_k\rangle\langle r_k|$, where $\hat\rho_0=\sum_k r_k |r_k\rangle\langle r_k|$ ($r_1\ge r_2\ge\cdots$) and $\hat H_0=\sum_k\epsilon_k |\epsilon_k\rangle\langle\epsilon_k|$ ($\epsilon_1\le\epsilon_2\le\cdots$). Writing $\hat U_I(\tau)=e^{i\hat H\tau}\hat U(\tau)\equiv e^{-i\hat\Lambda\tau}$ defines the Hermitian generator $\hat\Lambda=(i/\tau)\log_{\rm principal}\!\big(\hat U_I(\tau)\big)$.
Choosing a smooth interpolating function for the discharging rate, e.g. $\varphi(t)=\tau\sin^2(t\pi/(2\tau))$, we have
$\hat U_I(t)=e^{-i\hat\Lambda\varphi(t)}$, with $\varphi(\tau)=\tau$. The Hamiltonian $\hat H'(t)=\hat H_0+ \hat V(t)$ induces the desired discharge, where $\hat V(t)=\dot\varphi(t)e^{-i\hat H t}\hat\Lambda e^{i\hat H t}$ (for further details, see App.~\ref{Appendice1}).

\begin{figure}[t]
\centering
\includegraphics[width=\columnwidth]{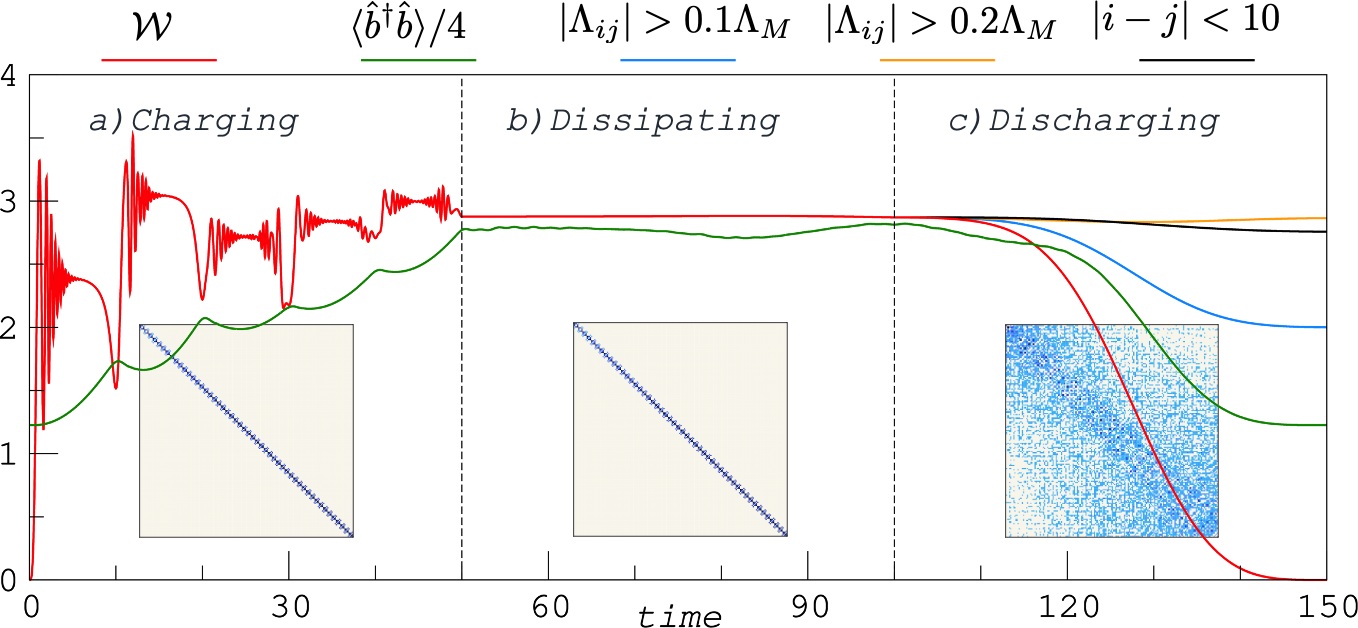}
\caption{Battery performance. 
(a)--(c) Ergotropy $\mathcal{W}$ (red curve) and average phonon occupation 
$\langle \hat b^\dagger \hat b\rangle/4$ (green curve) during charging, 
dissipation, and discharging. Parameters are as in Fig.~\ref{Fig_one}. 
The shaded maps in panels (a) and (b) show the corresponding battery Hamiltonian. 
In panel (c), the shaded map shows the Hermitian generator $\Lambda_{ij} =(\hat\Lambda)_{ij}$ associated 
with the exact discharge protocol of Ref.~\cite{Allahverdyan_2004}. The evolution 
of $\mathcal{W}$ is also shown for modified generators obtained by retaining only 
matrix elements with $|\Lambda_{ij}|/\Lambda_M>0.1$ (blue curve), 
$|\Lambda_{ij}|/\Lambda_M>0.2$ (orange curve), or $|i-j|<10$ (black curve), where 
$\Lambda_M=\max_{ij}|\Lambda_{ij}|$.}
\label{Fig_two}
\end{figure}

The resulting discharge dynamics correspond to the red and green portions of the curves in Fig.~\ref{Fig_two}c. The final value, $\mathcal{W}(t)\simeq10^{-7}$, indicates that essentially all the stored energy has been extracted from the battery. The two shaded maps in Fig.~\ref{Fig_two}c represent, respectively,  the matrix $\Lambda_{ij}=(\hat\Lambda)_{ij}$ and a magnified view of a portion of it. In contrast to $H_{HH}(t)$
whose matrix representation contains only near-diagonal couplings (shaded map in (a) and (b)), $\Lambda_{ij}$ is dense in the dimer--phonon number basis. Since the transformation from $\hat \Lambda$ to $\hat V(t)$ preserves this structure, implementing the perturbation $\hat V(t)$ proposed in Ref.~\cite{Allahverdyan_2004} would, in practice, require engineering a Hamiltonian containing operators of the form $b^n$ and $(b^\dagger)^n$, with $n$ a large positive integer (in principle up to $N_{ph}$), reflecting the presence of nonzero matrix elements far from the diagonal.

To assess how the discharge depends on the structure of $\hat{\Lambda}$, we prune
selected matrix elements $\Lambda_{ij}$ and recompute the evolution
[Fig.~\ref{Fig_two}(c)]. Removing elements with
$|\Lambda_{ij}|<0.2\Lambda_M$, or even $|\Lambda_{ij}|<0.1\Lambda_M$,
where $\Lambda_M=\max_{ij}|\Lambda_{ij}|$, significantly degrades the
discharge, highlighting the important role played even by small matrix
elements. As further diagnostic, we also remove matrix elements far from the diagonal, also finding
a strong reduction in the extracted work. These results show that both
the dense structure of $\hat{\Lambda}$ and its long-range off-diagonal couplings
are essential for optimal discharge, suggesting that the required $\hat{V(t)}$ is
so highly structured that its physical realization may be unrealistic.

The presence of phonons in the battery is particularly instructive in this context. The inverse-construction procedure already generates additional terms for a pure Hubbard dimer, but owing to its simplicity these may still appear physically plausible. By contrast, in the Hubbard--Holstein $V(t)$ contains couplings between widely separated phonon-number states, corresponding to high powers of the bosonic creation and annihilation operators. The phonons do not introduce the problem; they expose the unrealistic nature of the inverse-construction procedure in an unmistakable way.

\section{Discharging the battery: non-optimal protocol} \label{nonoptprot}
These considerations motivate the search for practically realizable discharge protocols. As a proof of concept, we consider a charged battery coupled at $t=0$ (the time origin is reset here) to a four-site, initially empty, non-interacting linear chain acting as the load. The load parameters are optimized
to maximize ergotropy ($\mathcal{W}$) extraction at a target time $t=T=10$. For the optimization we employ the Broyden–Fletcher–Goldfarb–Shanno (BFGS) algorithm \cite{Nocedal_Wright_2006} with analytical gradients of the time-evolution operator (see App.~\ref{Appendice2}).
Although based on a specific example, the following analysis allows us to draw some general conclusions.

\begin{figure}
\centering
\includegraphics[width=\columnwidth]{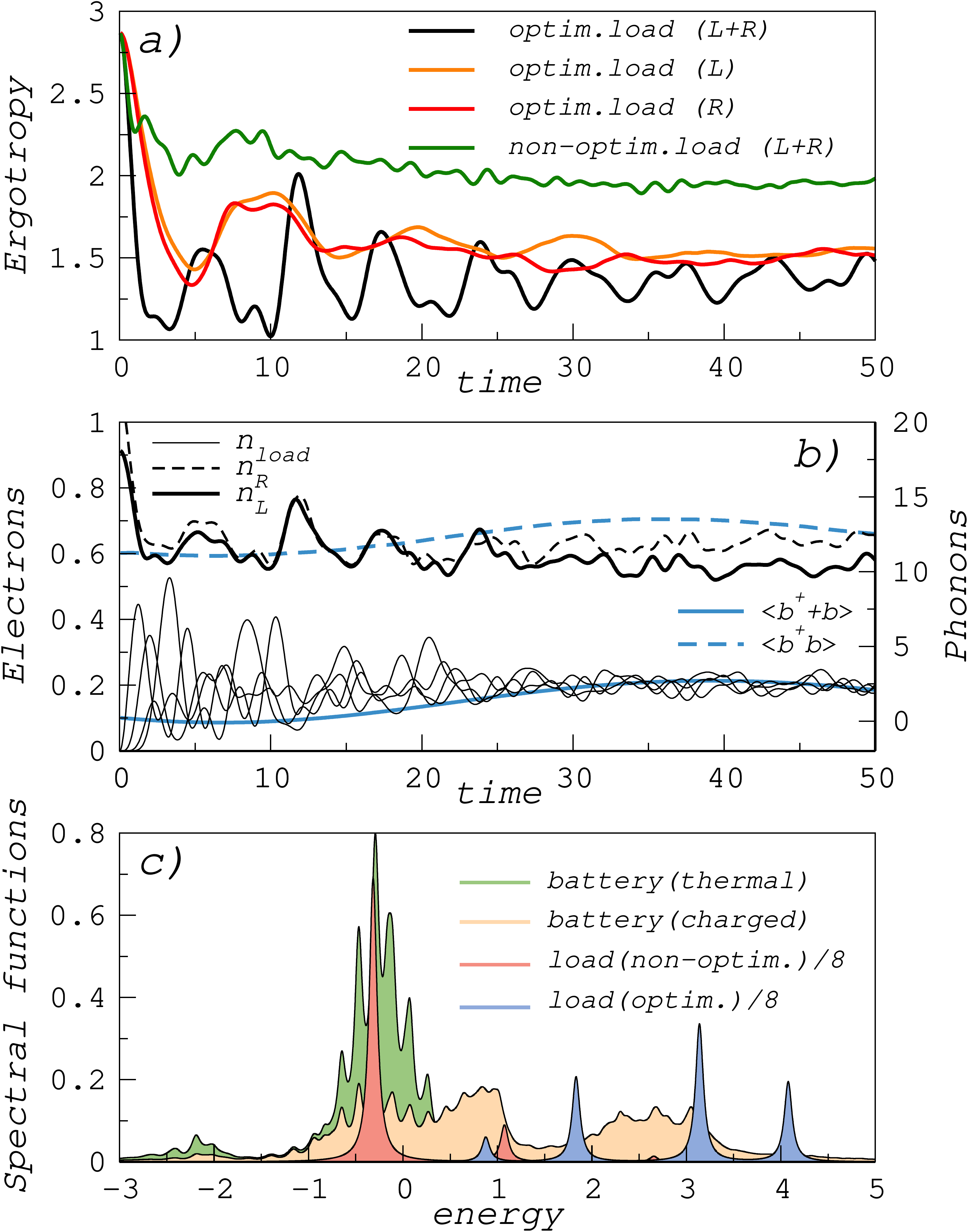}
\caption{(a) Ergotropy $\mathcal{W}$ during load optimization for four battery--load
couplings: optimized load coupled to both battery sites, to the left site, or
to the right site, and a non-optimized shifted load coupled to both sites.
(b) Optimized case: battery-site densities (thick black curves), load-site
densities (thin curves), and phonon displacement/occupation (blue curves).
Electronic and phononic quantities use the left and right axes, respectively.
(c) Thermal/charged battery spectra (green/orange shaded distributions) and
optimized/non-optimized load spectra (blue/red shaded distributions, scaled by $1/8$). 
For the removal spectra, the site averaged value
$[A^{(-)}_{\mathrm{batt}}(E;t;L)+A^{(-)}_{\mathrm{batt}}(E;t;R)]/2$ is shown.}
\label{Fig_three}
\end{figure}

Figure~\ref{Fig_three}a shows the reduction of $\mathcal{W}$ for four load configurations. The optimized battery--load connection yields by far the largest reduction, whereas coupling the load to a single battery site or perturbing the optimized load significantly (to different degrees) suppresses the discharge. 
During discharge, approximately 0.8 electrons are transferred to the load (Fig.~\ref{Fig_three}b). The electronic occupations closely track the ergotropy, whereas the phonon coordinate and average phonon number evolve on the slower phononic timescale. Partial revivals of $\mathcal{W}$ after $t\approx3$ originate from finite-size reflections. Similar behavior is found for chains with two, four, and six sites. Moreover, the optimized load extracts negligible energy from the battery in its initial thermal state, while a load optimized for the thermal state remains largely ineffective after charging.

Figure~\ref{Fig_three}c reveals the physical mechanism underlying these results. In analogy with tunnelling theory \cite{Bardeen61}, spectral functions characterize the phase space available for electron transfer and, hence, ergotropy extraction. Writing the battery density matrix at the end of charging ($t=\tau=100$) as $\rho_{\rm batt}(\tau)=\sum_{\lambda_2}p_{\lambda_2}(\tau)|\lambda_2\rangle\langle\lambda_2|$, where $|\lambda_2\rangle$ and $E_{\lambda_2}$ are the two-electron--phonon eigenstates and eigenvalues of the HH dimer, the relevant quantity is the electron-removal spectral function
\begingroup
\setlength{\abovedisplayskip}{2pt}
\setlength{\belowdisplayskip}{2pt}
\setlength{\abovedisplayshortskip}{2pt}
\setlength{\belowdisplayshortskip}{2pt}
\begin{align}
A^{(-)}_{\mathrm{batt}}(E;T;i)
&=\sum_{\lambda_2} p_{\lambda_2}(T)\times
\nonumber\\[-0.5ex]
&\sum_{n_1}
|\langle n_1|\hat c_{i\sigma}|\lambda_2\rangle|^2
\delta(E-E_{n_1}+E_{\lambda_2}),
\end{align}
\endgroup 
where $\hat c_{i\sigma}$ annihilates an electron of spin $\sigma$ at site $i=L,R$, while $|n_1\rangle$ and $E_{n_1}$ denote the one-electron eigenstates and eigenvalues of the dimer.

Compared to the initial state ($t=0$), the charged battery develops substantial high-energy spectral weight in its electron-removal spectral function (orange shaded curve), whereas the initial spectrum (green shaded curve) carries negligible weight for $E\gtrsim1$. It is worth noting in passing that a similar spectral distribution is also arrived at starting from a non-thermal state (see App.~\ref{Appendice4}). The electron-addition spectral function of the load at the contact site,
$A_\mathrm{load}^{(+)}(E)=\sum_{\mu_1}|\langle\mu_i|\hat c^\dagger_{1\sigma}|0\rangle|^2\delta(E-E_{\mu_1})$, 
shows that only the optimized load (blue shaded curve) has spectral weight in the same energy window; the non-optimized load (red shaded curve) lacks this overlap. Likewise, coupling the optimized load to the uncharged battery results in negligible particle transfer.

We also note that, beyond the spectral-matching argument relevant to the loading process, the charging-induced redistribution of the electronic spectrum may also provide indirect insight into the suppression of dissipation. It is plausible that an analogous redistribution occurs in the phonon-active excitations. For a finite-bandwidth bath, such a modification may reduce the overlap with the available dissipative channels, highlighting the role of finite bath bandwidth and spectral tailoring in limiting energy loss (see Appendix~\ref{Appendice5} for further discussion).

Overall, the results of Fig.~\ref{Fig_three} suggest a general design principle. Unlike conventional batteries, efficient work extraction is not universal but requires significant mutual spectral compatibility between a quantum battery and the load. 
A spectral function spanning a broad energy range increases the likelihood of such matching with finite loads. Increasing the load size may further facilitate matching through a denser spectrum, whereas interactions, disorder, and additional particles make it more stringent. This is where the relevance of the phonon degree of freedom comes into play: it generates many (vibronic) channels over a broad energy range, enabling efficient matching to the discrete spectra of finite loads, particularly small quantum ones, where spectral matching is intrinsically most demanding. Larger loads exhibit smaller level spacings, facilitating spectral matching. Interactions and/or additional particles, however, make the matching condition more stringent. More generally, beyond the specific system and optimization approach considered here, machine learning can substantially broaden the search for target loads for a given quantum battery. A graph neural network, trained on exact
Suzuki–Trotter time evolutions, enables an efficient exploration
of load topologies. Whereas the optimization performed here is restricted to continuous parameters within a fixed topology, ML can also optimize load topology and spectral overlap with the battery's spectral function. This provides a controlled way to determine whether the residual energy after optimal discharge is intrinsic to the battery or simply reflects the restricted design space, while identifying load topologies best matched to the battery's spectral function.

\section{Conclusions}  In summary, we have shown that a Hubbard--Holstein dimer can operate as a
correlation-driven quantum battery, with slow phonons providing a channel for
robust ergotropy storage. The model also
exposes a practical limitation: although complete discharge is possible in
principle through an optimal, one-size-fits-all unitary protocol, the required generator is highly
structured and unlikely to be physically realisable.

Efficient discharge therefore requires a load compatible with the microscopic
storage mechanism. In our finite-chain example, substantial work extraction is
obtained when the load spectrum matches the charged-battery spectrum and the
relevant transition matrix elements support the transfer. This suggests a
broader design logic: unlike classical batteries, whose operation is largely
set by macroscopic compatibility conditions, quantum batteries can be constrained
by the microscopic spectral and many-body structure of the load. Battery and
load should therefore be designed jointly, with the matching of excitation
spectra as a guide. Machine-learning-assisted searches could extend this
strategy to larger and more realistic nanosystems, where many-body effects in
the load play an important role. Work along these directions is currently under
way.
\appendix

\section{Suzuki-Trotter with Caldeira-Leggett bath}
\label{Appendice3}
In our system we have the following time-dependencies
\begin{align*}
 \hat{H}_{HH}&= \sum_{i\sigma} \epsilon_i (t)\hat{c}_{i\sigma}^\dagger \hat{c}_{i\sigma}
- J \sum_{\langle ij\rangle,\sigma} \hat{c}_{i\sigma}^\dagger \hat{c}_{j\sigma}
+ U \sum_i \hat{n}_{i\uparrow}\hat{n}_{i\downarrow} \nonumber \\
&\quad + \omega \hat{b}^\dagger \hat{b}
+ g(\hat{n}_1 - 1)(\hat{b}^\dagger+\hat{b}) +H_{\lambda}(t)\\
&= H_0 + H_\epsilon(t) + H_{\lambda}(t)
\end{align*}
The time-dependence of the Hamiltonians above factors $H_\epsilon(t) = f(t)H^0_\epsilon$ and $H_{\lambda}(t) = x(t)H^0_{\lambda}$. From now on the time-dependence of the Hamiltonians is implicit. Numerically, we write the propagator as
\begin{align*}
 U(t+\Delta,t) = e^{-i(H_0 +H_\epsilon(t+\Delta/2) + H_{\lambda}(t+\Delta/2))\Delta} + \mathcal{O}(\Delta^3)
\end{align*}
We now use the decomposition \cite{Suzuki}
\begin{align*}
 e^{-i\Delta (A+B)} = e^{-i\Delta B/2}e^{-i\Delta A}e^{-i\Delta B/2} + \mathcal{O}(\Delta^3)
\end{align*}
twice.
\begin{align*}
 U(t+\Delta,t) = e^{-i\Delta H_\epsilon/2} e^{-i(H_0 + H_{\lambda})\Delta}e^{-i\Delta H_\epsilon/2}  + \mathcal{O}(\Delta^3).
\end{align*}
Since $H^0_\epsilon$ is diagonal, we have $[e^{-i\Delta H_\epsilon/2}]_{mn} = \delta_{mn}e^{-i\Delta/2 [H^0_\epsilon]_{mm} f(t)}$; hence this part is numerically simple: 
\begin{align*}
 e^{-i(H_0 + H_{\lambda})\Delta} = e^{-iH_0 \Delta/2}\qty(e^{-iH^0_{\lambda}\Delta})^{x(t)}e^{-iH_0 \Delta/2} + \mathcal{O}(\Delta^3)
\end{align*}
Define $PDP^{-1}=e^{-iH^0_{\lambda}\Delta}$ 
\begin{align*}
 e^{-i(H_0 + H_{\lambda})\Delta} =e^{-iH_0 \Delta/2}PD^{x(t)}P^{-1}e^{-iH_0 \Delta/2} + \mathcal{O}(\Delta^3)
\end{align*}
Now the time-dependence is in the exponentiation of a diagonal matrix which is numerically simple. Hence, this gives a way to write the propagator without any need of diagonalization during the time-evolution \cite{Arvid26}.

\section{Prescription to find the perturbation  $V(t)$  
starting from the time evolution operator $\mathcal{U}$}
\label{Appendice1}

\noindent We describe in detail the procedure of Ref. \cite{Allahverdyan_2004} to find the external perturbation
that permits to fully discharge a battery. To start, we need a time evolution operator $U(t)$ such that $U(\tau)
\equiv \mathcal{U}$ , with $\mathcal{U}$  given (as in the article) by
\[ 
\mathcal{U}=\sum_k |\epsilon_k\rangle \langle r_k |,\textrm{ where } \rho_0=\sum_k r_k |r_k\rangle \langle r_k |, ~ H =\sum_{k'} \epsilon_{k'} |\epsilon_{k'} \rangle \langle \epsilon_{k'} |, 
\]
and $\epsilon_1 \le \epsilon_2 \le \epsilon_3 ..., ~ r_1 \ge r_2 \ge r_3 ...$ Consider the discharging final time $\tau$: there, the time-evolution operator in the interaction picture  looks like $U_I(\tau)= e^{iH \tau}\mathcal{U}$. $U_I(\tau)$ is unitary (to check, simply multiply $\mathcal{U}$ and $\mathcal{U}^\dagger$).
We must find the Hermitian generator $\Lambda$ so that $ U_I(\tau)= e^{-i\Lambda\tau} $; to this end we take the matrix logarithm, i.e.  $\Lambda = \frac{i}{\tau}\,\log_{\it principal}\!\big(U_I(\tau)\big)$; this is done via diagonalization  as follows:\\

\noindent i) Diagonalize in the usual way $U_I(\tau)$, for example looking at the right eigenvectors (the matrix is not hermitian):
$U_I(\tau)\,|\phi_k\rangle = m_k\,|\phi_k\rangle,
\qquad |m_k|=1 \qquad \textrm {($U_I(\tau)$ is unitary)}$\\

\noindent ii) Write each eigenvalue as a phase
$ m_k = e^{-i\lambda_k \tau}$. However, as usually done with unitary matrices, 
we must consistently chose all the  eigenvalues in the same logarithmic branch.
The standard choice is to go for the
 \emph{principal branch} and  impose
$
\mathrm{Arg}(m_k)\in(-\pi,\pi]
\;\;\Rightarrow\;\;
\lambda_k \in \left(-\frac{\pi}{\tau},\,\frac{\pi}{\tau}\right].$\\

\noindent iii) Then
$
\Lambda
=  \frac{i}{\tau}\,\log_{\text{principal}}\!\big(U_I(\tau)\big) = \sum_k \lambda_k\,|\phi_k\rangle\langle\phi_k|
$.\medskip

\noindent Choose a smooth interpolation function $\varphi(t)$ which connects $0$ and $\tau$ and 
satisfies $\varphi(0)=
\dot\varphi(0)=
\dot\varphi(\tau)=0, 
\varphi(\tau)=\tau.$ In the paper, $\varphi(t)=\tau \sin^2(\frac{\pi t}{2\tau})$: As interpolation, the interaction-picture propagator at all times $t\in [0,\tau]$  is 
\[
U_I(t)= e^{-i\Lambda\,\varphi(t)} .
\]
Then in the interaction-picture, the equation for $U_I(t)$ is
$i\dot U_I(t)=V_I(t)U_I(t)$, giving $V_I(t)= \dot\phi(t)\,\Lambda$ (verified 
by substitution. By transforming back to Schr\"odinger picture, we get
\[
V(t)= \dot\phi(t)\, e^{-iH_0 t}\,\Lambda\, e^{iH_0 t}.
\]
So the driving Hamiltonian to discharge the system between 0 and $\tau$ is
$H'(t)=H+V(t)$ and  $U(\tau)=U.$
This is the procedure we implemented numerically to produce the results
in Fig. 2c of the main paper.

\begin{figure}
\centering
\includegraphics[width=0.2\textwidth]{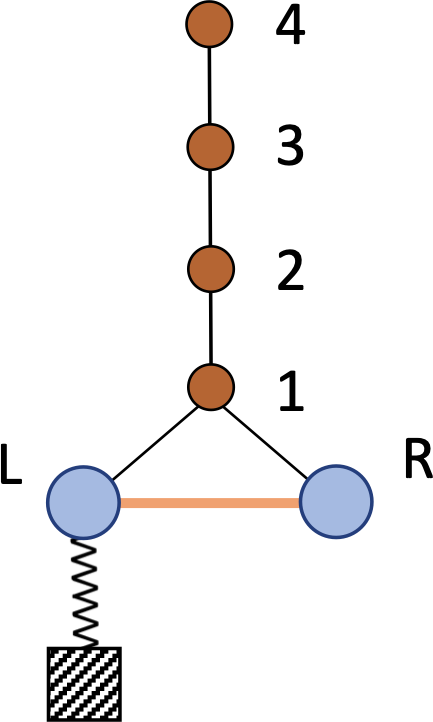}
\caption{Site labels for the battery+chain system}
\label{Fig_chain}
\end{figure}
\section{Minimization}
\label{Appendice2}
Consider the battery coupled to the external chain $H_{ch}$
\begin{align}
H &= H_{\rm HH}+H_{\rm ch} \nonumber\\
&= \sum_{i\sigma}\epsilon_i(t)\hat c_{i\sigma}^\dagger \hat c_{i\sigma}
-J\sum_{\langle ij\rangle,\sigma}\hat c_{i\sigma}^\dagger \hat c_{j\sigma}
+U\sum_i \hat n_{i\uparrow}\hat n_{i\downarrow} \nonumber\\
&\quad +\omega \hat b^\dagger \hat b
+g(\hat n_1-1)\hat x
+\sum_{i\sigma}V_i(t)\left(\hat c_{i\sigma}^\dagger \hat d_{1\sigma}+{\rm H.c.}\right) \nonumber\\
&\quad +\sum_{\langle ij\rangle\sigma}V_{ij}\hat d_{i\sigma}^\dagger \hat d_{j\sigma}
+\sum_{i\sigma}\varepsilon_i \hat d_{i\sigma}^\dagger \hat d_{i\sigma}.
\end{align}
Here $\varepsilon_i$ denotes the on-site energies of the external chain and
$V_{ij}$ its nearest-neighbor hopping amplitudes. We consider two possible
couplings $V_i$ between the battery sites and the first site of the external
chain (Fig.~\ref{Fig_chain}). The protocol therefore contains $2L+1$ free parameters. At $t=\tau=100$, the dissipation is switched off and the battery is coupled to the
chain, $V_i(t)=\theta(t-\tau)V_i$. The battery is then discharged from $t=\tau$ to $t=t_d=110$, however, we let the evolution continue to $t=150$. We define the battery energy as $E(t)=\Tr[H_{HH}\Tr_{ch}\rho(t)] = \Tr[H_{HH}\Tr_{ch}U(t,\tau)\rho(\tau) U(t,\tau)^\dagger]$, and then we minimize $E(t_d)$ with respect to $V_i$, $V_{ij}$ and $\varepsilon_i$. We employ the Broyden–Fletcher–Goldfarb–Shanno (BFGS) algorithm \cite{Nocedal_Wright_2006} for the optimization, taking advantage of analytically computed gradients to accelerate convergence \cite{WilcoxMath}. For any parameter $p=V_i,V_{ij},\varepsilon_i$ we have the derivative:
\begin{align*}
 \dv{U(t,\tau)}{p} &= \dv{p}e^{-iH(p)(t-\tau)} =\dv{p} e^{-iH(p)T} \\
 &=(-i) \int_0^T e^{-iH (T-u)}\pdv{H}{p} e^{-uH} \dd u\\
 &=(-i) \sum_{\lambda \tilde{\lambda}}\ket{\lambda}\int_0^T e^{-i\lambda (T-u)}\mel{\lambda}{\pdv{H}{p}}{\tilde{\lambda}} e^{-i\tilde{\lambda}u} \bra{\smash{\tilde{\lambda}}} \dd u\\
 &=(-i) \sum_{\lambda \tilde{\lambda}} \ket{\lambda}\mel{\lambda}{\pdv{H}{p}}{\tilde{\lambda}}  \bra{\smash{\tilde{\lambda}}} e^{-i\lambda T} \int_0^T e^{-i(\tilde{\lambda}-\lambda)u} \dd u\\
 &= \sum_{\lambda \tilde{\lambda}} \ket{\lambda}\mel{\lambda}{\pdv{H}{p}}{\tilde{\lambda}}  \bra{\smash{\tilde{\lambda}}} e^{-i\lambda T}  \frac{1}{\tilde{\lambda}-\lambda}\eval{e^{-i(\tilde{\lambda}-\lambda)u}}_{0}^T\\
 &= \sum_{\lambda \tilde{\lambda}} \ket{\lambda}\mel{\lambda}{\pdv{H}{p}}{\tilde{\lambda}}  \bra{\smash{\tilde{\lambda}}}   \frac{1}{\tilde{\lambda}-\lambda} \qty[e^{-i\tilde{\lambda} T} - e^{-i\lambda T}] 
\end{align*}
For the 4-site chain we find the parameters:\\ 
$\varepsilon_1 =2.51514403,\varepsilon_2 =2.49358016, \varepsilon_3 =  2.48404278 , \varepsilon_4 =2.43343502$\\
$V_1 = 0.68718053$, $V_2=-0.57459215$\\
$V_{12}=-0.88139746$, $V_{23}=-0.89688703$, $V_{34}=-1.1856921$\\
For the non-optimized parameters the only change we do is that we set $\varepsilon_1=0$. Similarly, for the load at the right site we set $V_1=0$ and for the load at the left site we set $V_2=0$.
This minimisation process is stopped when the absolute value of all components of the gradient are less than $10^{-5}$.\\
For the battery we calculate the removal part of the spectral function, i.e. we do not include the part which corresponds to particle addition:
\begin{align*}
A^{(-)}_{j}(E,t)= \sum_{ \lambda; i} p_{\lambda}(t) |\langle\phi_i| c_{j\sigma} |\psi_\lambda\rangle|^2 \delta(   (\lambda-E_i) -E ),
\end{align*}
where $\psi_\lambda$ are the two-particle eigenstates of the isolated battery with eigenvalue $\lambda$, i.e. eigenstates of the initial Hamiltonian. The $p_{\lambda}(t)$ are weights of each eigenstate as a function of time. Initially they are $p_{\lambda}=\frac{e^{-\lambda \beta}}{\sum_\lambda e^{-\lambda \beta}}$ with $\beta=2$. The states $\phi_i$ are the the one-electron eigenstates (in the presence of phonons) of the isolated dimer with eigenvalue $E_i$\\
We also calculate the local density states (LDOS) of the first site in the external chain:
\begin{align*}
 D_{11}(\omega)=\sum_\lambda |\langle\lambda|1\rangle|^2 \delta(  \lambda-\omega).
\end{align*}
The spectral function and the LDOS are plotted via Lorentzians which are broadened with $\gamma= 0.05$.
\section{Charging  from a non-thermal state}
\label{Appendice4}
We have found that the charged state is easily achievable even if one starts from quite a different state. We consider, as in \cite{Emil21,Arvid26},
\begin{align}
 \rho=e^{-H_{ph}\beta}/\Tr(e^{-H_{ph}\beta}) \otimes \qty(c^\dagger_{1\uparrow}c^\dagger_{2\downarrow}\ketbra{0}c_{2\downarrow}c_{1\uparrow}) \label{eq:non_thermal},
\end{align}
where only the phonon is now in the heat bath. 
\begin{figure}[H]
\centering
\includegraphics[width=0.8\textwidth]{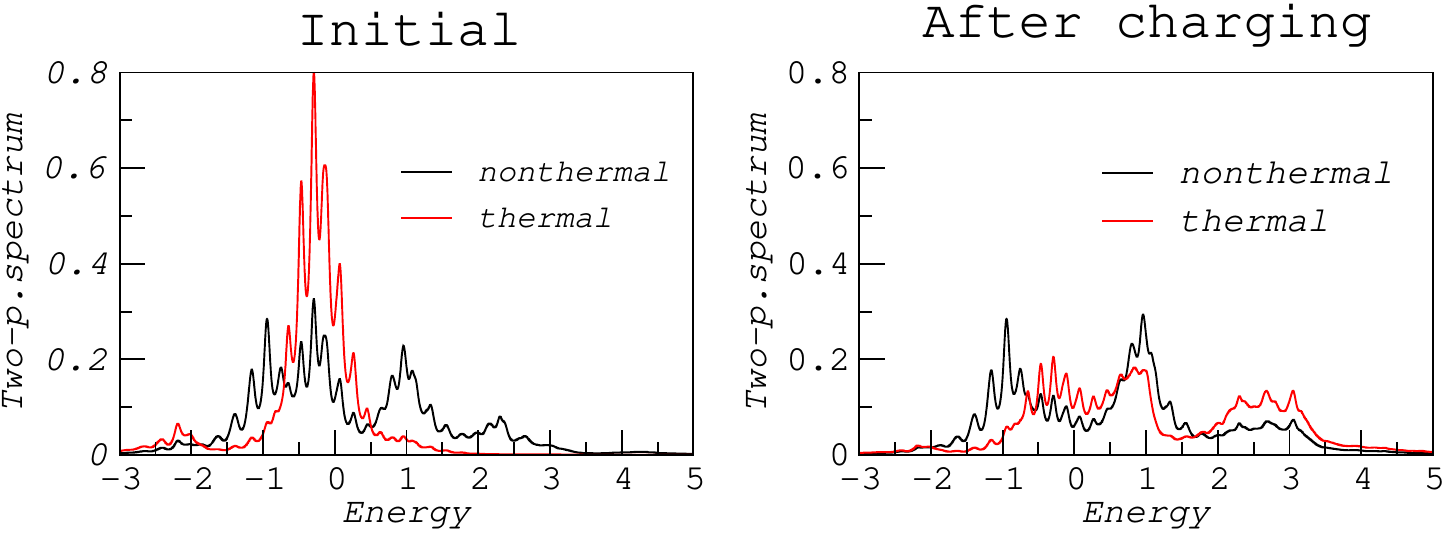}
\caption{Comparison between the spectral functions for the thermal state considered in the main text and the spectral functions obtained from \eqref{eq:non_thermal}}
\label{Fig_APP4}
\end{figure}
The initial distributions at $t=0$ are quite different  (Fig.~\ref{Fig_APP4},left), since the initial states are different. After the charging process we find that also the non-thermal state has weight in the region which is relevant for discharging (Fig.~\ref{Fig_APP4},right). 

\section{Caldeira--Leggett bath and finite bandwidth}
\label{Appendice5}
In our model, dissipation is governed not only by the system--bath coupling matrix elements, but also by the spectral overlap between the bath and the relevant transitions of the interacting system. The Caldeira--Leggett (CL) bath we have chosen has finite spectral support, with oscillator frequencies restricted to the interval ($\omega\in[0,1]$). To place this observation in context, we resort to an heuristic argument based on the electron-removal spectrum $A^{(-)}_{\rm batt.}$ shown in Fig.~\ref{Fig_three}(c).

The spectrum $A^{(-)}_{\rm batt.}$ does not directly represent the transitions sampled by the phonon bath, since the latter couples through the phonon displacement operator and conserves the electronic particle number. Nevertheless, the spectral redistribution in $A^{(-)}_{\rm batt.}$ provides a qualitative indication that charging substantially modifies the characteristic excitation energies of the coupled electron--phonon system. At equilibrium, the spectral weight obtained from the full finite-temperature density matrix is concentrated predominantly outside the frequency window supported by the CL bath, as shown by the green shaded curve in Fig.~\ref{Fig_three}(c) of the main text. During charging, the site-dependent potential further redistributes the spectral weight. At the end of the protocol, a substantial fraction of the electron-removal spectrum lies outside the bath bandwidth, as shown by the orange shaded curve in Fig.~\ref{Fig_three}(c) of the main text. This redistribution is therefore consistent with a reduction in the relaxation channels accessible to the bath and may help explain the ability of the battery to retain a significant fraction of the stored energy. 

This interpretation remains qualitative: a conclusive assessment of the spectral picture would require the population- and matrix-element-weighted spectral function associated with the phonon displacement operator.
Still from a spectral perspective, one could also argue that our protocol simply places the charged system in a frequency regime where the bath couples inefficiently.
Yet, the case of the non-optimized load in Fig.~\ref{Fig_three} suggests a subtler picture: although the electronic addition and removal spectra are misaligned, energy still dissipates, albeit more weakly than in the optimized case.
Indeed, by analogy with the non-optimized electronic case, some dissipation might also be expected in the phononic case, yet very little is observed. Spectral mismatch and matrix-element weighting may therefore be insufficient to provide a complete explanation, as the driving protocol can generate nonequilibrium populations and correlations that are not captured by the corresponding equilibrium spectral functions. This dynamical aspect becomes apparent at the end of the protocol:
the system parameters return to their initial values, so that, if allowed to equilibrate, the system would relax toward the corresponding equilibrium state. Nevertheless, the phononic mode remains in an excited and displaced configuration, indicating that the suppression of dissipation is also a dynamical consequence of the charging protocol.
As already anticipated in connection with the spectral interpretation, a rigorous determination of the relative roles of spectral mismatch and protocol-induced dynamics would require not only a direct calculation of the phononic spectrum, but also a full characterization of the nonequilibrium response generated by the driving protocol; we defer both analyses to future work.

As a final remark, we note that the opposite assumption of a broadband environment can also be physically restrictive. In particular, phenomenological Markovian Lindblad descriptions based on frequency-independent local dissipators effectively assign relaxation channels over a broad range of transition frequencies, thereby neglecting the finite or structured spectral support characteristic of realistic environments. More microscopic Lindblad constructions can instead retain frequency-dependent rates evaluated at the Bohr frequencies of the system and therefore need not suffer from this limitation. From this perspective, suppressing relaxation by shifting the relevant excitations into regions of low environmental spectral density may be regarded as a form of spectral engineering rather than as an artificial suppression of dissipation.

\funding{E.Ö and C.V. gratefully acknowledge financial support from the Swedish Research Council (Vetenskapsrådet, VR, Grant No. 2017-03945 and No. 2022-04486).}

\suppdata{The data that support the findings of this study are available on request
from the corresponding author. The data are not publicly available due to
privacy or ethical restrictions.}

\bibliographystyle{iopart-num}
\bibliography{bib1}

\end{document}